# G–BAM: A Generalized Bandwidth Allocation Model for IP/MPLS/DS-TE Networks

Rafael Freitas Reale, Romildo Martins da S. Bezerra
DMCC / GSORT
Federal University of Bahia / Federal Institute of Bahia
Salvador, Bahia, Brazil
reale@ifba.edu.br, romildo@ifba.edu.br

Joberto S. B. Martins
NUPERC
Salvador University
Salvador, Bahia, Brazil
joberto@unifacs.br

*Abstract* — Bandwidth Allocation Models (BAMs) configure and handle resource allocation (bandwidth, LSPs, fiber) in networks in general (IP/MPLS/DS-TE, optical domain, other). BAMs currently available for IP/MPLS/DS-TE networks (MAM, RDM, G-RDM and AllocTC-Sharing) basically define resource restrictions (bandwidth) by "class" (traffic class, application´s class, user´s class or other grouping criteria) and allocate on demand this resource. There is a BAM allocation policy inherent for each existing model which behaves differently under distinct network state, such as heavy traffic loads and dynamic traffic and/or application scenarios. A generalized Bandwidth Allocation Model (G-BAM) is proposed in this paper. G-BAM, firstly, incorporates the inherent behavior of currently used BAMs such as MAM, RDM, G-RDM and AllocTC-Sharing in IP/MPLS/DS-TE context. G-BAM, secondly, proposes a new policy/ behavior allocation in addition to existing ones in which additional private resources are incorporated. G-BAM, thirdly, allows a smoother BAM policy transition among existing policy alternatives resulting from MAM, RDM and AllocTC-Sharing adoption independently. The paper focuses on the first characteristics of G-BAM which is to reproduce MAM, RDM and AllocTC-Sharing behaviors. As such, the required configuration to achieve MAM, RDM and AllocTC-Sharing behaviors is presented followed by a proof of concept. Authors argue that the G-BAM reproducibility characteristics may improve overall network resource utilization under distinct traffic profiles.

## I. Introduction and Motivation

Bandwidth Allocation Models (BAMs) configure and handle resource allocation (bandwidth, LSPs, fibre) in networks in general (IP/MPLS/DS-TE, optical domain, other). In IP/MPLS/DS-TE networks, Bandwidth Allocation Models are used with the main objective of define rules and limits for link utilization by defining Bandwidth Constraints (BCs) for traffic classes (TCs) [3][6]. In practice, these models effectively define how bandwidth resources are obtained and shared among applications and/or users.

The main proposed BAMs for IP/MPLS/DS-TE networks like MAM – Maximum Allocation Model, RDM – Russian Dolls Model and AllocTC-Sharing have distinct operational characteristics that finally result in resource optimization for a specific traffic profile [12]. In general, each proposed BAMs treat distinct traffic profiles with a different behavior and, as such, an expected resource optimization might be compromised in case the network traffic profile changes and does not match the inherent optimization behavior of the installed BAM. The Table I resumes BAM´s resource allocation for distinct network traffic profiles [8].

TABLE I – BAMs BEHAVIORAL AND OPERATIONAL CHARACTERISTICS (MAM, RDM AND ALLOCTC-SHARING)

| BAM – Behavioral Characteristics | MAM | RDM | AllocTC-Sharing |
|---|---|---|---|
| Link utilization with a traffic profile composed by a large amount of low priority traffic | Low | High | High |
| Link utilization with a traffic profile composed by a large amount of high priority traffic | Low | High | High |
| TCs isolation | High | Medium | Low |
| **BAM Operational Characteristics** | **MAM** | **RDM** | **AllocTC-Sharing** |
| "High-to-Low" (HTL) sharing | No | Yes | Yes |
| "Low-to-High" (LTH) sharing | No | No | Yes |

In brief, the MAM model targets network traffic profiles in which a strong isolation between traffic classes (TCs) is required [4]. In this model, TCs use only private resources and there is no bandwidth (resource) sharing among TCs (Figure 01).

RDM and AllocTC-Sharing have the main objective of maximize link utilization and, in order to achieve this goal, they allow resource sharing among traffic classes (TCs). This inherent behavior reduces traffic isolation between traffic classes.

The RDM, in brief, allows the sharing of non-allocated resources belonging to high priority traffic classes by low priority traffic classes (HTL loan or HTL-TCx behavior - Figure 01) [5]. This model tends to improve link utilization for a network traffic profile with a large volume of low priority TCs and/or applications. In addition, it inherently tends to create a larger volume of LSP setups for low priority TCs [8].

The AllocTC-Sharing model keeps RDM resource allocation strategy of "High-To-Low" loans and adds the possibility of "Low-To-High" loans (LTH-TCx behavior – Figure 01). As such, AllocTC-Sharing allows low priority classes (TCs) to get resources normally reserved for high priority classes (TCs). In brief, "loans" are allowed in both directions (HTL e LTH). This model targets networks in which link utilization is expected to be maximized with a weak isolation among TCs being acceptable. This corresponds, typically, to networks with high priority elastic applications like multimedia services, among others [10].

BAMs presented till now, have an excluding behavior between each other. In MAM model, all resources are private and no sharing is allowed. RDM has no private resource and all resource configured for high priority TCs might be got by low priority TCs. AllocTC-Sharing has no private resource as well and allows "loans" in both directions (HTL and LTH).

The G-RDM model was a first attempt to have a more flexible bandwidth allocation model [2]. G-RDM is a hybrid model in which the "HTL loan" strategy of RDM incorporates the private resource strategy defined by MAM. Overall G-RDM operation results in having a configurable volume of private resources with the remaining resources having the capability to be loaned by lower priority traffic classes (TCs).

As discussed till now, there is a BAM allocation policy inherent for each existing model (MAM, RDM, G-RDM and AllocTC-Sharing) which behaves differently under distinct network profile/state, such as heavy traffic loads and dynamic traffic and/or application scenarios. A second implication of this brief introduction to BAMs is that, to our knowledge, there is no solution that effectively integrates private resources with existing BAM models and with "LTH loans" strategy.

Considering this fact, a Generalized Bandwidth Allocation Model (G-BAM) is proposed in order to allow more flexibility in terms of resource allocation strategies for networks. An additional motivation for G-BAM is to allow and facilitate the adoption of the best suitable BAM model in accordance with the dynamics of network traffic profiles.

The following sections of this paper will present initially a brief background in relation to DS-TE networks that is the target network type (IP/MPLS/DS-TE) considered in the paper for G-BAM. Following that, G-BAM model is explained and formally introduced. Finally, G-BAM capability to reproduce MAM, RDM and AllocTC-Sharing models presented by indicating the required dynamic configuration and the behavior equivalence in relation to the BAM models is validated using simulation as a proof of concept.

## II. DS-TE Architecture - Background

IETF (Internet Engineering Task Force) has proposed in [6] the requirements for DS-TE (DiffServ aware Traffic Engineering) networks with the basic objective to exploit the benefits MPLS-TE (MPLS Traffic Engineering) [1] and DiffServ (Differentiated Services Architecture) [11] technologies. DS-TE architecture, among other possible applications, supports traffic engineering. In this context, Traffic Classes (TCs) group LSPs (Label Switched Path) for sets of applications and define the resources that can be allocated for these LSPs setup as required dynamically by applications. As such, DS-TE may be understood fundamentally as being a management model supporting the availability of bandwidth for users (Applications/ LSPs).

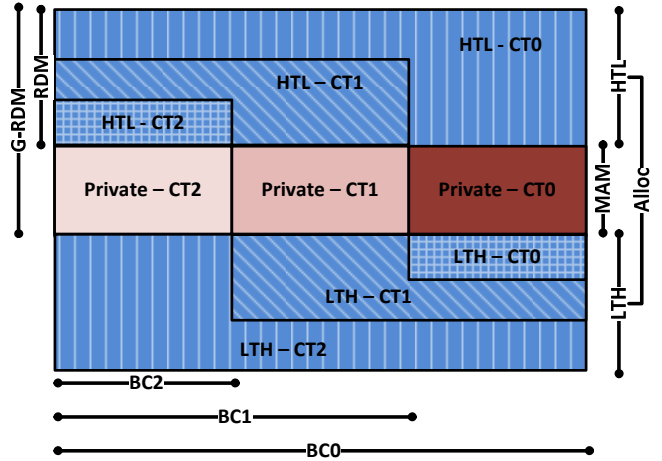

Fig. 1. BAMs and resource allocation strategies

DS-TE makes MPLS aware of Traffic Class (TC), applications and, indirectly, their SLA (Service Level Agreements) and required Quality of Service (QoS). In practice, TC definition, organizes and defines priorities for a variety of applications (multimedia, web, others) existing in a multi-service network. DS-TE defines 08 different Traffic Classes (TCs) (TC0-TC7) and each TC may accommodate any set of applications and services defined by network administration, management, traffic engineering computation or DiffServ implementation.

In DS-TE, by convention, TC0 is the class supporting best effort traffic (lower priority) and TC7 hosts the higher priority traffic. Obviously, each TC may accommodate a number of applications (services) mapped on any number of label switched paths (LSPs) transporting application's traffic.

## III. G-BAM – A Generalized Bandwidth Allocation Model - Description

The Generalized Bandwidth Allocation Model (G-BAM) is a new bandwidth allocation model proposed that integrates in a configurable way various resource (bandwidth) allocation strategies such as private TCs resources and loans between high and low priority classes/ applications. A brief description of the configurable resource allocation possibilities follows:

- Private Resource (Private) – the configured resource is private (unique) to a specific Traffic Class (TC);

- Loan "high to low" (high-to-low - HTL) – in this configurable allocation method, the resource (bandwidth) allocated to higher priority CTs that are not being currently used may be borrowed by lower priority CTs (LSPs); and

- Loan "low to high" (low-to-high - LTH) – in this configurable allocation method, the resource (bandwidth) allocated to lower priority TCs that are not being currently used may be borrowed for higher priority TCs (LSPs).

In G-BAM, it can be defined (configured) for each traffic class: (i) the amount of private resources; (ii) the HTL loan limit; and (iii) the LTH loan limit. The overall operation results in a bandwidth allocation model that includes the main models (MAM, RDM, AllocTC-Sharing and G-RDM) in a single generalized scheme. Beyond that, and not less important, G-BAM still allows new intermediate configuration settings between existing models open new perspectives in terms of applications and user support for network management in this specific context of resource allocation [9].

## IV. G-BAM Description

G-BAM model can be described as follows:

1. $BC_i$ is the Bandwidth Constraint for Traffic Class "i" ($TC_i$). The total amount of configured bandwidth constraints should not exceed the maximum available bandwidth for the link (M):

$$\sum_{i=0}^{C-1} BC_i \leq M$$

2. For each defined $TC_i$, a maximum allowed loan "High to Low" ($HTL_i$) and "Low to High" ($LTH_i$) is defined. The $HTL_i$ and $LTH_i$ values should not exceed the configured $BC_i$:

$$HTL_i \leq BC_i \; e \; LTH_i \leq BC_i$$

3. The private bandwidth for each $TC_i$ is obtained as follows:

$$PRIVATE_i = (BC_i - \max(HTL_i, LTH_i))$$

4. $N_i$ is the total bandwidth allocated to LSPs belonging to traffic class "i". The maximum value for $N_i$ is as follows:

$$\max(N_i) \leq BC_i + \sum_{j=i+1}^{C-1} HTL_j + \sum_{k=0}^{i-1} LTH_k$$

5. The available bandwidth for loan ($HTL_{DISPi}$, $LTH_{DISPi}$) in a $TC_i$ is defined as follows:

$$HTL_{DISPi} = \min\{HTL_i, [BC_i - \max(PRIVATE_i, N_i)]\}$$

$$LTH_{DISPi} = \min\{LTH_i, [BC_i - \max(PRIVATE_i, N_i)]\}$$

6. $N_i$, at execution time, is restricted by the available loans (low and high) as follows:

$$N_i \leq BC_i + \sum_{j=i+1}^{C-1} HTL_{DISPj} + \sum_{k=0}^{i-1} LTH_{DISPk}$$

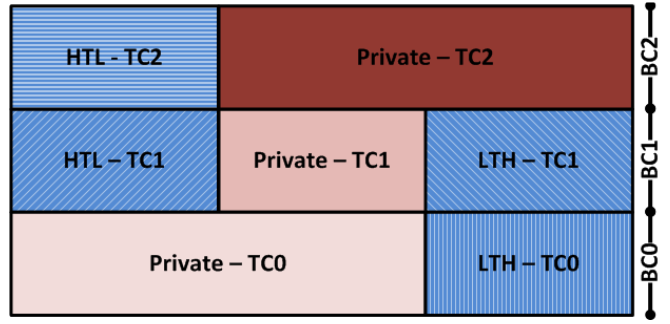

Fig. 2. G-BAM Model

## V. Case Study – G-BAM Reproducing the behavior of multiple BAMs

In this section, a case study of G-BAM will be shown. The main purpose of the case study is to demonstrate the configurable characteristics of G-BAM in reproducing MAM, RDM and AllocTC-Sharing BAM behaviors. The general configuration parameters of the G-BAM for this case study are:

- Links: 622 Mbps (STM-4 - SDH)
- Traffic Classes (TCs): TC0, TC1 and CTC2
- Restrictions bandwidth: In Table II.

TABLE II. Bandwidth Allocation by Traffic Class (CT)

| BC | Max BC (%) | MAX BC ( Mbps) |
|---|---|---|
| BC0 | 40 | 248,80 |
| BC1 | 35 | 217,70 |
| BC2 | 25 | 155,50 |

### A. CASE STUDY 01 – G-BAM Configured to Reproduce MAM Behavior

In order to get a MAM behavior out of G-BAM is only necessary to configure HTL and LTH loans to 0% (Table III). As such, G-BAM will consider that the total amount of resources (bandwidth) destined to TC (BC) will be private (Section III – Rule 3).

TABLE III – G-BAM Configuration for Reproduce MAM Behavior

| HTL (%) | HTL (Mbps) | LTH (%) | LTH (Mbps) | Private |
|---|---|---|---|---|
| 0% | 0 | 0% | 0 | 248,80 |
| 0% | 0 | 0% | 0 | 217,70 |
| 0% | 0 | 0% | 0 | 155,50 |

In this configuration (Fig. 3), there are no loans and, as occurs in MAM operation, only private resources limited by BCs are allocated for TCs applications.

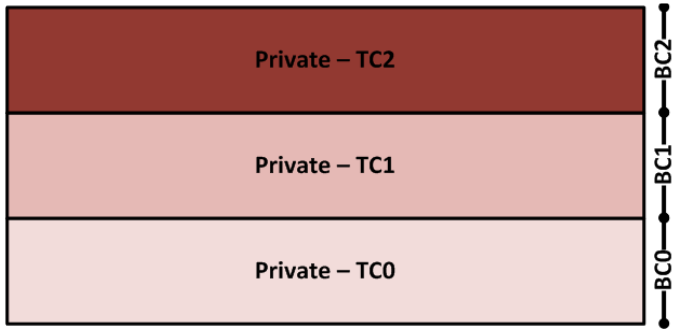

Fig. 3. Case Study 01 – Resulting Configuration – MAM Behavior.

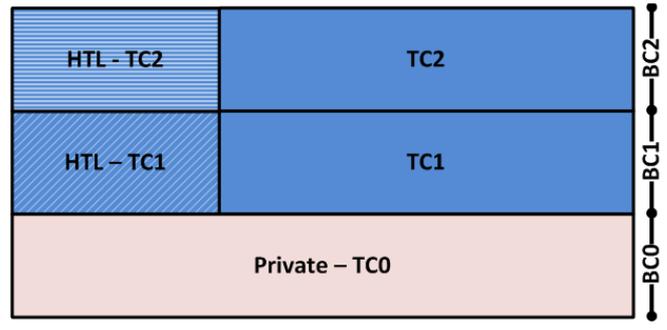

Fig. 4. Case Study 02 - Resulting Configuration – RDM Behavior.

In Table IV we observe that the maximum bandwidth to be used by LSPs belonging to a TC (MAX (Ni)) is composed only by its bandwidth restriction.

TABLE IV – G-BAM REPRODUCING MAM BEHAVIOR (MAXIMUM BANDWIDTH BY TC).

|  | BC | HTL | | | | LTH | | | | Total |
|---|---|---|---|---|---|---|---|---|---|---|
|  |  | CT0 | CT1 | CT2 | Total | CT0 | CT1 | CT2 | Total | MAX (Ni) |
| CT0 | 248,80 | 0 | 0 | 0 | 0 | 0 | 0 | 0 | 0 | 248,8 |
| CT1 | 217,70 | 0 | 0 | 0 | 0 | 0 | 0 | 0 | 0 | 217,7 |
| CT2 | 155,50 | 0 | 0 | 0 | 0 | 0 | 0 | 0 | 0 | 155,5 |

*B. CASE STUDY 02 – G-BAM Configured to Reproduce RDM Behavior*

In order to get a RDM behavior as a result of G-BAM configuration it is necessary to configure HTL loans to 100% (Table V). As such, G-BAM will consider that the total amount of resources (bandwidth) allocated to higher priority CTs, and which are not being used, can be loaned by lower priority TCs LSPs.

TABLE V – G-BAM CONFIGURATION REPRODUCING RDM BEHAVIOR.

| HTL (%) | HTL (Mbps) | LTH (%) | LTH (Mbps) | Private |
|---|---|---|---|---|
| 0% | 0 | 0% | 0 | 248,80 |
| 100% | 217,7 | 0% | 0 | 0 |
| 100% | 155,5 | 0% | 0 | 0 |

The resulting behavior of this configuration, like I RDM, is that resources not being used by lower priority classes are allocated for applications belonging to higher priority classes (Fig. 4).

Table VI presents the amount of bandwidth per TC and, as result of rule IV (section III), the maximum bandwidth of a TC is composed by its bandwidth constraint (BC) plus the bandwidth loaned by higher TCs.

TABLE VI – RDM REPRODUCED BY G-BAM (BANDWIDTH PER TC).

|  | BC | HTL | | | | LTH | | | | Total |
|---|---|---|---|---|---|---|---|---|---|---|
|  |  | CT0 | CT1 | CT2 | Total | CT0 | CT1 | CT2 | Total | MAX (Ni) |
| CT0 | 248,8 | 0 | 217,7 | 155,5 | 373,2 | 0 | 0 | 0 | 0 | 622,0 |
| CT1 | 217,7 | 0 | 0 | 155,5 | 155,5 | 0 | 0 | 0 | 0 | 373,2 |
| CT2 | 155,5 | 0 | 0 | 0 | 0 | 0 | 0 | 0 | 0 | 155,5 |

*C. CASE STUDY 03 – G-BAM Configured to Reproduce AllocTC-Sharing Behavior*

The AllocTC-Sharing behavior can be obtained from G-BAM model by setting the HTL and LTH loans to 100% (Table VII). Thus, G-BAM considers that the whole bandwidth allocated to higher priority TCs, which are not being used, can be loaned for LSPs of lower priority TCs and vice-versa.

TABLE VII – G-BAM CONFIGURATION REPRODUCING ALLOCTC-SHARING BEHAVIOR

| HTL (%) | HTL (Mbps) | LTH (%) | LTH (Mbps) | Private |
|---|---|---|---|---|
| 0% | 0 | 100% | 248,8 | 0 |
| 100% | 217,7 | 100% | 217,7 | 0 |
| 100% | 155,5 | 0% | 0 | 0 |

As in AllocTC-Sharing behavior, G-BAM allows loans in both directions (LTH and HTL) for all defined TCs. This effectively corresponds to the expected AllocTC-Sharing behavior as illustrated in Figure 5.

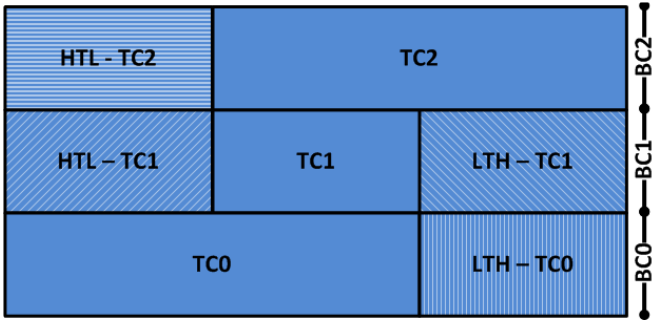

Fig. 5. Case Study 03 – Resulting Configuration – AllocTC-Sharing Behavior.

As a consequence of Section III – Rule 4, it is observed in Table VIII that the maximum bandwidth to be used by LSPs belonging to a TC is composed by its bandwidth constraint (BC) plus the bandwidth loaned of higher TCs and also by the bandwidth loaned of lower TCs.

TABLE VIII – AllocTC-Sharing Behavior Reproduced by G-BAM (bandwidth per TC).

|  | BC | HTL | | | | LTH | | | | Total MAX ($N_i$) |
|---|---|---|---|---|---|---|---|---|---|---|
|  |  | CT0 | CT1 | CT2 | Total | CT0 | CT1 | CT2 | Total |  |
| CT0 | 248,8 | 0 | 217,7 | 155,5 | 373,2 | 0 | 0 | 0 | 0 | 622 |
| CT1 | 217,7 | 0 | 0 | 155,5 | 155,5 | 248,8 | 0 | 0 | 248,8 | 622 |
| CT2 | 155,5 | 0 | 0 | 0 | 0 | 248,8 | 217,7 | 0 | 466,5 | 622 |

## VI. G-BAM Simulation Reproducing MAM, RDM e AllocTC-Sharing Behavior

In the previous section, G-BAM assumed distinct configurations that, intuitively, indicate that it can reproduce the behavior of current BAM models MAM, RDM and AllocTC-Sharing. In this section, we complement the case study presented with a proof of concept by simulating G-BAM using a simple point-to-point link topology and comparing the results with MAM, RDM and AllocTC-Sharing simulations using the same topology. The main objective is to compare behaviors of the different G-BAM configurations presented with MAM, RDM and AllocTC-Sharing models. It is important to mention that the potential flexibility and dynamic behavior of G-BAM is not the target of the presented simulations that is focused of validating the reproducibility characteristics of G-BAM model.

The simulated topology (Fig. 6) uses 01 traffic sources (S1), one destination (D) and two scenarios are evaluated.

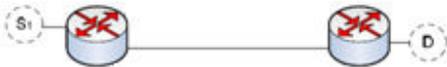

Fig. 6. – Proof-of-Concept – Simulated Topology

The configuration parameters of the validation scenarios are as follows:

- Link: 622 Mbps (STM-4 – SDH)
- Existing Traffic Classes: TC0, TC1 and TC2
- Bandwidth Constraints, according to Table IV

TABLE IV - Bandwidth Constraint (BCs) per Traffic Class (TCs)

| BC | Max BC (%) | MAX BC ( Mbps) |
|---|---|---|
| BC0 | 40 | 248,80 |
| BC1 | 35 | 217,70 |
| BC2 | 25 | 155,50 |

This evaluation uses the bandwidth allocation model simulator named BAMSim (Bandwidth Allocation Model Simulator) developed in [13] and based in [7].

The evaluation scenarios are as follows:

- Scenario 01: traffic generated is initially higher for TCs of higher priority and then gets higher for all classes.
- Scenario 02: traffic generated in initially higher for TCs of lower priority and then gets higher for all classes.

The first scenario has the purpose to demonstrate that G-BAM has equivalent behavior to MAM and RDM models. The simulation scenario enforces the high-to-low loan strategy that is inherent to RDM. Concerning MAM, the private resources allocation is verified.

The second scenario objective is to demonstrate that G-BAM has equivalent behavior to AllocTC-Sharing. This is verified by enforcing the low-to-high (LTH) allocation strategy used by AllocTC-Sharing.

### A. Scenario 01 - Description and Results Evaluation

In this simulation scenario the parameter "link load by TC" is evaluated. The simulation run parameters are as follows:

- Number of LSPs – 1.000
- Evenly distributed LSP bandwidth: 5 to 10 Mbps
- Exponential modeled "$S_1$" LSP request arrival intervals as follows:
  - LSPs – $TC_0$ – 3 s
  - LSPs – $TC_1$ - 3 s - delay of 800 s
  - LSPs – $TC_2$ - 3 s - delay of 1400 s
- Exponentially modeled LSP time life: average of 250 seconds (should cause link saturation)
- Simulation stop criteria: number of LSPs

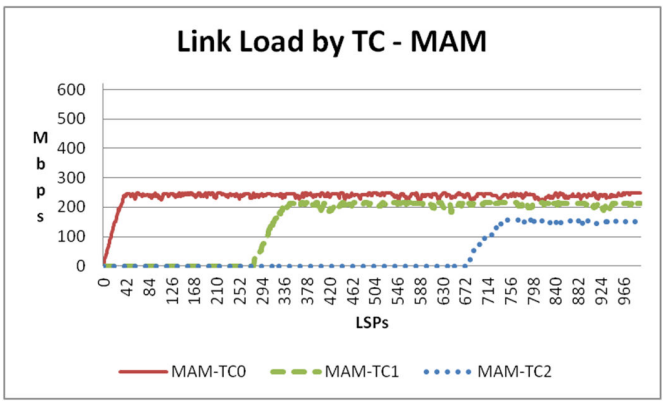

Fig. 7. Link Load by TC - MAM

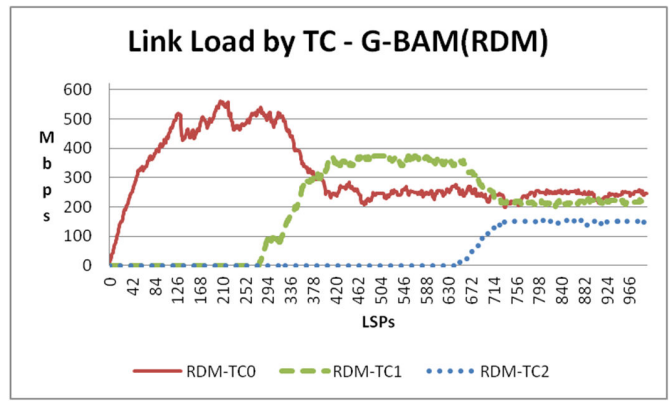

Fig. 10. Link Load by TC - G-BAM (Configured to have RDM Behavior)

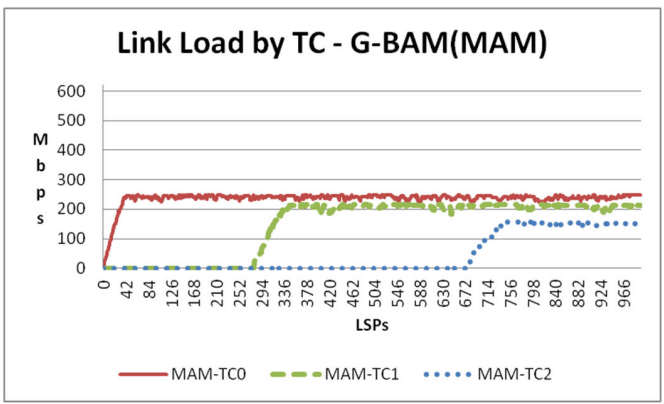

Fig. 8. Link Load by TC -G-BAM (Configured to have MAM Behavior)

The simulation runs shows that G-BAM effective reproduces the behavior of MAM, RDM and AllocTC-Sharing by adopting distinct parameter configuration. Figures 07 and 08 demonstrate the G-BAM behavior equivalence for MAM with its private resource allocation model.

Figures 09 and 10 demonstrate G-BAM a behavior equivalent to RDM allowing HTL loans.

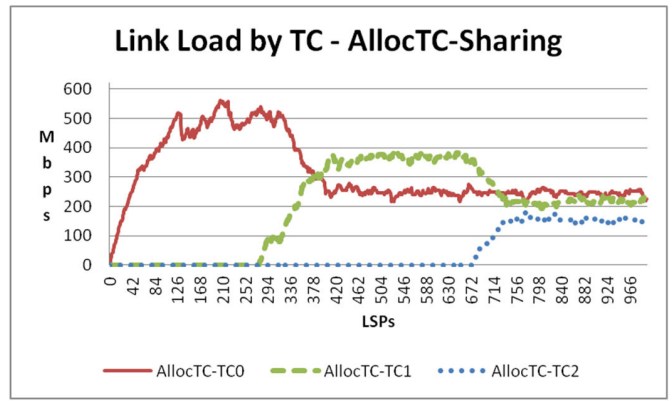

Fig. 11. Link Load by TC - AllocTC-Sharing

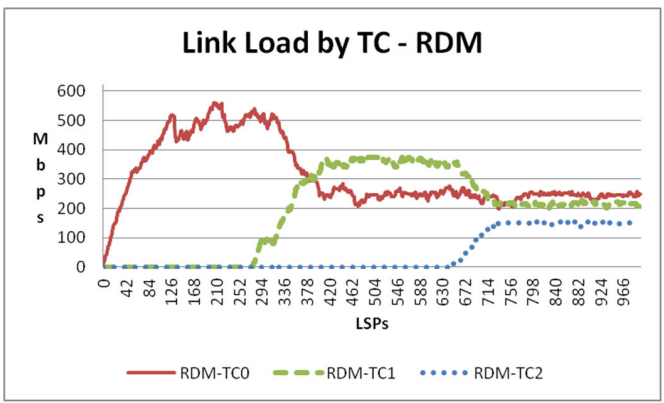

Fig. 9. Link Load by TC -RDM

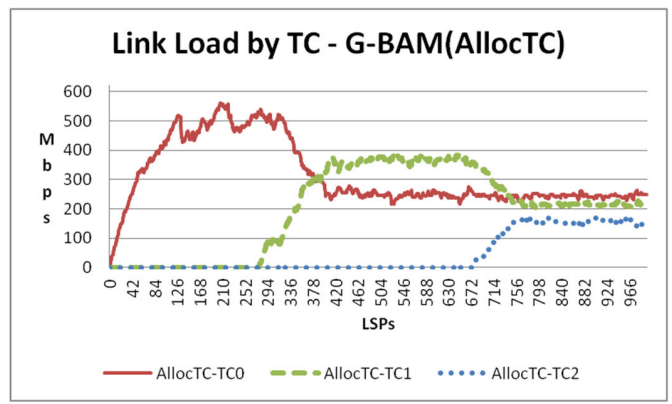

Fig. 12. Link Load by TC - G-BAM (Configured to have AllocTC-Sharing Behavior)

Figures 11 and 12 demonstrate the G-BAM behavior equivalence for AllocTC-Sharing allowing HTL and LTH loans depending on the traffic profile currently supported by the network.

## B. Scenario 02 - Description and Results Evaluation

In this simulation scenario the parameter "link load by TC" was evaluated with a distinct traffic profile. The simulation run parameters are as follows:
- Number of LSPs – 1.000
- Evenly distributed LSP bandwidth: 5 to 10 Mbps
- Exponential modeled "$S_1$" LSP request arrival intervals as follows:
  - LSPs – $TC_0$ – 3 s - delay of 1400 s
  - LSPs – $TC_1$ - 3 s - delay of 800 s
  - LSPs – $TC_2$ - 3 s
- Exponentially modeled LSP time life: average of 250 seconds (should cause link saturation)
- Simulation stop criteria: number of LSPs

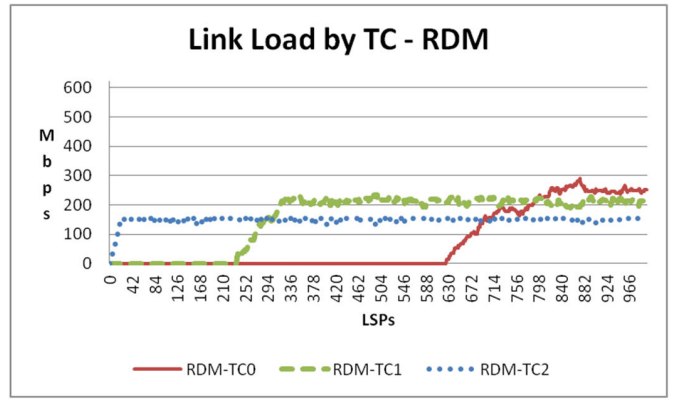

Fig. 15. Link Load by TC - RDM

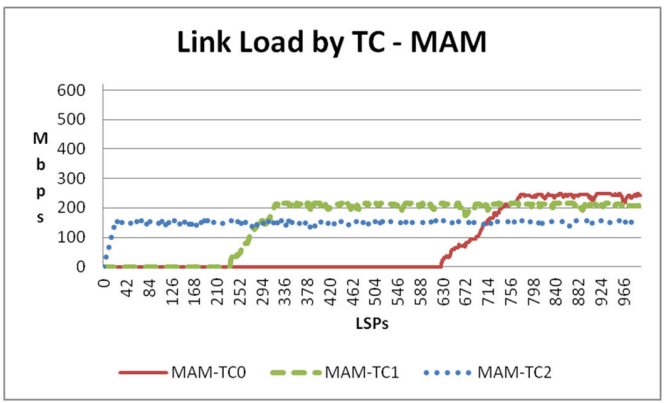

Fig. 13. Link Load by TC - MAM

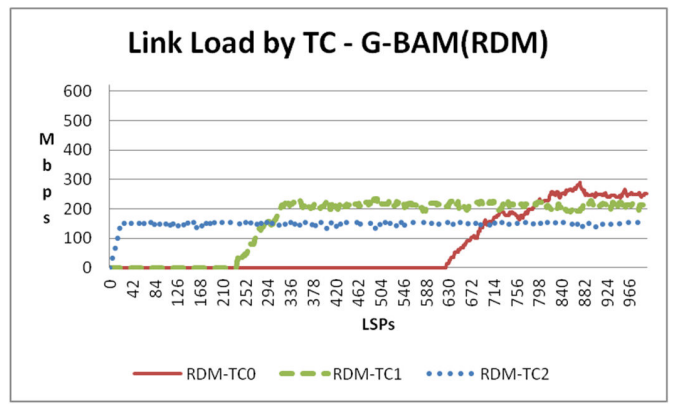

Fig. 16. Link Load by TC - G-BAM (Configured as RDM)

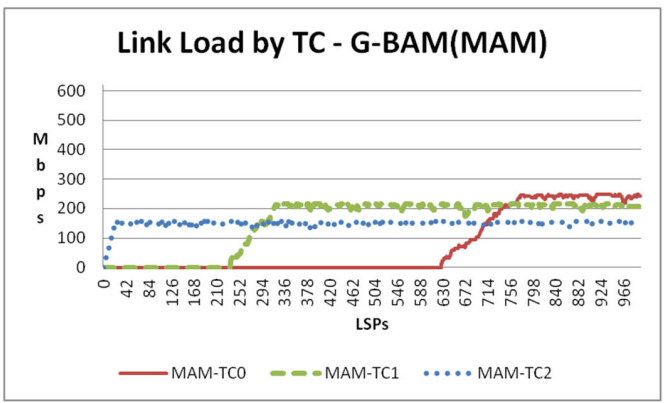

Fig. 14. Link Load by TC - G-BAM (Configured as MAM)

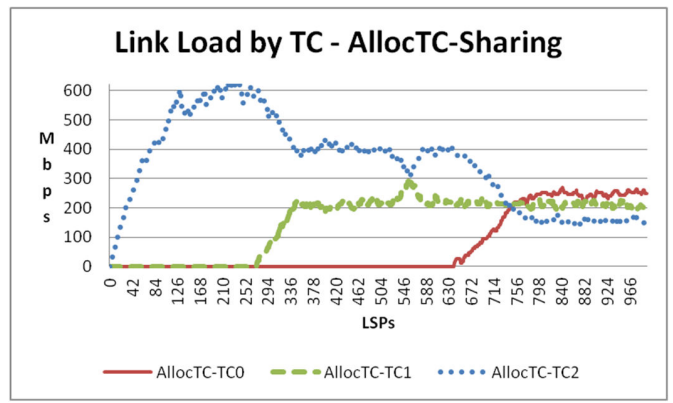

Fig. 17. Link Load by TC - AllocTC-Sharing

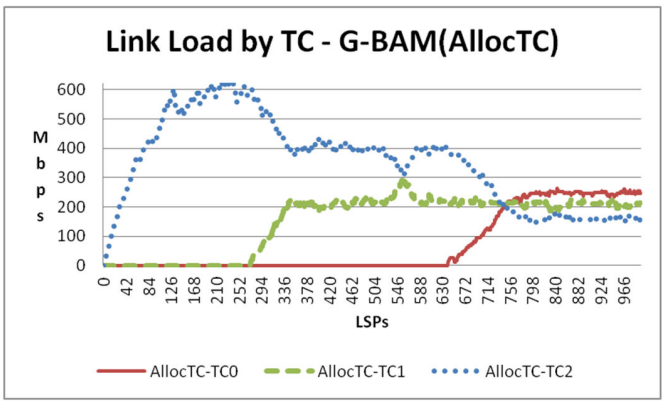

Fig. 18. Link Load by TC - G-BAM (Configured as AllocTC-Sharing)

This second simulation runs also shows that G-BAM effectively reproduces the behavior of MAM, RDM and AllocTC-Sharing for a traffic profile initially saturated with high priority TCs applications. Figures 13 and 14 again demonstrate the G-BAM behavior equivalence for MAM which is equivalent to previous simulation scenario making use only of private resources. Figures 15 and 16 demonstrate that G-BAM has behavior equivalence for RDM allowing HTL loans in a scenario where loans occur less frequently due to saturation of lower priority traffic. Figures 17 and 18 demonstrate the G-BAM behavior equivalence with AllocTC-Sharing allowing HTL and LTH loans under traffic condition with higher priority traffic being saturated at the beginning of the simulation.

## VII. FINAL CONSIDERATIONS

The Generalized Bandwidth Allocation Model (G-BAM) has the capability to reproduce the behavior of current available BAMs such as MAM, RDM and AllocTC-Sharing in a single model and, as such, generalizes the inherent behavior of these BAMs in a single implementation.

An important advantage of adopting G-BAM instead of a single BAM model instance in a network is to provide network managers with a single solution (model) that allows the maximization of network and link utilization with diverse traffic profiles. In effect, G-BAM provides some kind of "adaptability" since it may be configured to have distinct behavior for distinct traffic profiles.

As a matter of fact, without G-BAM there are other two possibilities achieving a similar result as indicated above in terms of the maximization of network and link utilization:

- Firstly, network managers may evaluate or infer the network traffic profile and adopt the current BAM (MAM, RDM or AllocTC-Sharing) that fits better the assumed network traffic profile. The drawback of this approach is that the network traffic profile is assumed inherently to be static. As such, any modification of network traffic profile currently supported by the network may reduce or even compromise network and link utilization as a whole.

- A second possibility (evaluated in [9]) is to adopt a set of BAMs (either MAM, RDM or AllocTC-Sharing) and provide a switching scheme between them whenever the network traffic profile changes. This approach get a similar result as indicated for G-BAM but still has a problem of handling preemptions and loans resulting from a BAM model that does not suits anymore the current traffic profile. This operational aspect of BAM switching has been discussed in [9].

Another G-BAM inherent advantage that has not been totally explored in this paper is that, since it is a single model, the rules for preemption and loans may be adjusted to provide a smooth migration among the behavior of current existing BAMs. In fact, G-BAM may potentially cope with the dynamics of the network traffic profile and have sets of configured behaviors for them, including transition patterns of behaviors. This approach is not possible with distinct BAM implementations due to the distinct policies of their inherent associated traffic profiles.

Beyond all that, G-BAM also incorporates new resource allocation strategy. In effect, it is now possible with G-BAM to define private resources and HTL and LTH loans for all traffic classes. New allocation strategies include:

- The integration of private resources with LTH loans; and

- The integration of private resources with the behavior of AllocTC-Sharing.

This provides a set of additional capabilities that might be capable to support new classes of traffic profiles that have not been supported till now with MAM, RDM and AllocTC-Sharing in a single or multi-BAM implementation.

As future work it is planned to explore these new allocation capabilities provided by G-BAM, evaluate G-BAM dynamics in relation to network traffic profile and, finally, to evaluate new preemption strategies that might facilitate the management of multiservice networks.